\documentclass[english,10pt,twocolumn,twoside]{IEEEtran}
\usepackage[T1]{fontenc}
\usepackage[latin9]{inputenc}
\usepackage{bm}
\usepackage{amsmath}
\usepackage{amsthm}
\usepackage{amssymb}

\makeatletter
\theoremstyle{plain}
\newtheorem{thm}{\protect\theoremname}
\theoremstyle{plain}
\newtheorem{prop}[thm]{\protect\propositionname}

\usepackage{colortbl}
 \usepackage{cite}

\makeatother

\usepackage{babel}
\providecommand{\propositionname}{Proposition}
\providecommand{\theoremname}{Theorem}

\begin{document}

\title{On the Maximal Invariant Statistic for Adaptive Radar Detection in
Partially-Homogeneous Disturbance with Persymmetric Covariance}

\author{D.~Ciuonzo,\IEEEmembership{~Member,~IEEE,} D. Orlando,\IEEEmembership{~Senior~Member,~IEEE,}
and L.~Pallotta,\IEEEmembership{~Member,~IEEE} \thanks{Manuscript received 26th August 2016.\protect \\
D. Ciuonzo and L. Pallotta are with University of Naples \textquotedbl{}Federico
II\textquotedbl{}, DIETI, Via Claudio 21, 80125 Naples, Italy. E-mail:
domenico.ciuonzo@ieee.org; luca.pallotta@unina.it. \protect \\
D. Orlando is with Università degli Studi \textquotedblleft Niccolò
Cusano\textquotedblright , Via Don Carlo Gnocchi 3, 00166 Roma, Italy.
E-mail: danilo.orlando@unicusano.it.\protect \\
}}
\maketitle
\begin{abstract}
This letter deals with the problem of adaptive signal detection in
partially-homogeneous and persymmetric Gaussian disturbance within
the framework of invariance theory. First, a suitable group of transformations
leaving the problem invariant is introduced and the Maximal Invariant
Statistic (MIS) is derived. Then, it is shown that the (Two-step)
Generalized-Likelihood Ratio test, Rao and Wald tests can be all expressed
in terms of the MIS, thus proving that they all ensure a Constant
False-Alarm Rate (CFAR).\end{abstract}


\section{Introduction}

\IEEEPARstart{A}{daptive detection} has attracted enormous interest
in the last decades (see e.g. \cite{Gini2001} and references therein).
Most design solutions rely on the Homogeneous Environment (HE), wherein
a set of secondary data (ideally free of useful signal) is available,
sharing the same spectral properties of the disturbance as in the
cell under test (primary data) \cite{Kelly1986,Fuhrmann1992}. Though
the HE often leads to elegant closed-form solutions ensuring satisfactory
performance \cite{Ciuonzo2015,Liu2014}, relevant scenarios are often
non-homogeneous due to environmental factors and system considerations
\cite{McDonald2000,Melvin2000,Ward1994}. Frequently a non-homogeneous
scenario is depicted through the Partially-Homogeneous Environment
(PHE, subsuming the HE as a special case), i.e., both the test data
and secondary data share the same covariance up to an unknown scaling
factor\footnote{Indeed, while keeping a relative tractability of the model, this assumption
provides increased robustness to power level fluctuations of the disturbance
between the test cell and the set of training data, which may manifest
for instance due to variations in terrain and the use of guard cells
\cite{McDonald2000,Ward1994,Wicks2006}.}. 

The Adaptive Normalized Matched Filter (ANMF) \cite{Conte1994,Conte1996}
(or Adaptive Coherence Estimator (ACE) \cite{Scharf1996}) is the
most common detector employed in PHE. In fact, it corresponds to the
Generalized Likelihood Ratio Test (GLRT) for the aforementioned model,
as shown in \cite{Kraut1999}, and to a two-step GLRT (2S-GLRT) design
procedure\footnote{This is tantamount to devising the GLRT under known covariance of
the disturbance and then making it adaptive via its substitution with
the sample covariance matrix of secondary data.}. More recently, literature has shown that the ANMF also corresponds
to Rao and Wald tests \cite{A.DeMaio2008}.

Interestingly, other studies concerning detectors design in PHE appeared
in the literature in the last years. These include works exploiting
the peculiarity of the PHE at the design stage, along with the assumption
of a \emph{persymmetric covariance} structure, as adopted in devising
a plain GLRT in \cite{Casillo2007}. The latter structure arises when
a sensing system is equipped with a symmetrically-spaced linear array
(or uses symmetrically-spaced pulse trains) and it is thus able to
collect statistically symmetric data in forward/reverse directions\footnote{We remark that the persymmetric property is not only limited to linear
arrays, but it also arises in different geometries such as standard
rectangular arrays, uniform cylindrical arrays (with an even number
of elements), and some standard exagonal arrays \cite{VanTrees2002}. }. Such constraint introduces dependences among the unknowns of the
disturbance and can be exploited \emph{to reduce the number of secondary
data needed for adaptive processors.} More recently, Rao and Wald
tests were derived according to the same philosophy in \cite{Hao2012a}.
Similarly, a persymmetric version of ACE (referred to as ``Per-ACE'')
was obtained in \cite{Gao2014}, as the result of a 2S-GLRT technique
in a PHE.

All the aforementioned works differ from \cite{A.DeMaio2015}, where
the adaptive detection problem has been handled by resorting to the
\emph{Principle of Invariance} \cite{Scharf1991,Lehmann2006}. Note
however that \cite{A.DeMaio2015} is restricted to the HE assumption.
Indeed, when exploited at the design stage, the principle of invariance
allows to focus on decision rules enjoying some desirable (practical)
features. The first step consists in identifying a suitable group
of transformations which leaves unaltered: ($i$) the formal structure
of the hypothesis testing problem, ($ii$) the data distribution family
and ($iii$) the useful signal subspace. Of course, the group invariance
requirement leads to a (lossy) data reduction. The least compression
of the original data ensuring the desired invariance is represented
by\emph{ }the\emph{ Maximal Invariant Statistic} (MIS), which organizes
the original data into equivalence classes. Hence, every invariant
test can be expressed in terms of the MIS \cite{Lehmann2006}. Consequently,
the parameter space is usually compressed after reduction by invariance
and the dependence on the original parameters set is mapped into the
so-called\emph{ induced maximal invariant} \cite{Lehmann2006}. Referring
to Radar adaptive detection, the mentioned principle represents the
workhorse for obtaining a statistic which is invariant with respect
to (w.r.t.) the set of nuisance parameters, thus constituting the enabler
for CFAR detectors.

The contributions of this letter are summarized as follows. Unlike
\cite{A.DeMaio2015}, the Principle of Invariance is exploited to
obtain the MIS and, hence, invariant architectures, assuming the PHE
and persymmetric covariance. Specifically, following the lead of \cite{A.DeMaio2015},
the problem at hand is recast in canonical form, which facilitates
the derivation of the MIS and allows to gain insights on the problem.
Then, the group of transformations which leaves the problem invariant
is identified and the explicit expression of the MIS w.r.t. the above
transformation group is derived. Remarkably, closed-form expressions
for the (2S-) GLRT, the Rao test, and the Wald test \cite{Kay1998}
are derived and shown to be all function of the data solely through
the MIS, thus proving their CFARness\footnote{\emph{Notation} - Lower-case (resp. Upper-case) bold letters denote
vectors (resp. matrices), with $a_{n}$ (resp. $A_{n,m}$) representing
the $n$-th (resp. the $(n,m)$-th) element of the vector $\bm{a}$
(resp. matrix $\bm{A}$); $\mathbb{R}^{N}$, $\mathbb{C}^{N}$, and
$\mathbb{H}^{N\times N}$ (resp. $\mathbb{S}^{N\times N}$) are the
sets of $N$-dimensional vectors of real numbers, of complex numbers,
and of $N\times N$ Hermitian (resp. symmetric) matrices, respectively,
while $\mathbb{R}^{+}$ denotes the set of positive-valued real numbers;
$\mathbb{E}\{\cdot\}$, $(\cdot)^{T}$, $(\cdot)^{\dagger}$, $\mathrm{Tr}\left[\cdot\right]$,
$\left\Vert \cdot\right\Vert $, $\Re\{\cdot\}$ and $\Im\{\cdot\}$,
denote expectation, transpose, Hermitian, matrix trace, Euclidean
norm, real part, and imaginary part operators, respectively; $\bm{0}_{N\times M}$
(resp. $\bm{I}_{N}$) denotes the $N\times M$ null (resp. identity)
matrix; $\bm{0}_{N}$ (resp. $\bm{1}_{N}$) denotes the null (resp.
ones) column vector of length $N$; $\det(\bm{A})$ denotes the determinant
of matrix $\bm{A}$; $\bm{A}\otimes\bm{B}$ indicates the Kronecker
product between matrices $\bm{A}$ and $\bm{B}$; the symbol ``$\sim$''
means ``distributed as''; $\bm{x}\sim\mathcal{C}\mathcal{N}_{N}(\bm{\mu},\bm{\Sigma})$
denotes a complex (proper) Gaussian-distributed vector $\bm{x}$ with
mean vector $\bm{\mu}\in\mathbb{C}^{N\times1}$ and covariance matrix
$\bm{\Sigma}\in\mathbb{H}^{N\times N}$; $\bm{X}\sim\mathcal{C}\mathcal{N}_{N\times M}(\bm{A},\bm{B},\bm{C})$
denotes a complex (proper) Gaussian-distributed matrix $\bm{X}$ with
mean $\bm{A}\in\mathbb{C}^{N\times M}$ and $\mathrm{Cov}[\mathrm{vec}(\bm{X})]=\bm{B}\otimes\bm{C}$.}.

The rest of the letter is organized as follows: in Sec.~\ref{sec: Problem formulation},
we formulate the problem under investigation; in Sec.~\ref{sec: MIS},
we obtain the MIS for the problem at hand and show invariance for
the considered detectors; finally, in Sec.~\ref{sec: Conclusions}
we provide some concluding remarks. Proofs are confined to the Appendix
and to a supplemental material document.

\section{Problem Formulation \label{sec: Problem formulation}}

In this section, we describe the detection problem at hand and recall
its canonical form representation \cite{A.DeMaio2015}. Assume that
a sensing system collects data from $N>1$ (spatial and/or temporal)
channels. The returns from the cell under test, after pre-processing,
are properly sampled and arranged in $\bm{r}\in\mathbb{C}^{N\times1}$.
We want to test whether $\bm{r}$ contains useful target echoes or
not. Additionally, we assume that a set of secondary (signal-free)
data, $\bm{r}_{k}\in\mathbb{C}^{N\times1}$, $k=1,\ldots K$ (with
$K\geq2N$), is available. In summary, the decision problem at hand
can be formulated in terms of the following binary hypothesis test
\begin{equation}
\begin{cases}
\mathcal{H}_{0}: & \begin{cases}
\bm{r}=\bm{n}_{0},\\
\bm{r}_{k}=\bm{n}_{0k},\quad k=1,\ldots,K,
\end{cases}\\
\mathcal{H}_{1}: & \begin{cases}
\bm{r}=\alpha\,\bm{s}+\bm{n}\\
\bm{r}_{k}=\bm{n}_{0k},\quad k=1,\ldots,K,
\end{cases}
\end{cases}\label{eq: Original hypothesis testing problem}
\end{equation}
where
\begin{itemize}
\item $\bm{s}\in\mathbb{C}^{N\times1}$ is the nominal steering vector ($\left\Vert \bm{s}\right\Vert =1$),
exhibiting a persymmetric structure, that is, $\bm{s}=\bm{J}\bm{s}^{*}$
with $\bm{J}\in\mathbb{R}^{N\times N}$ a suitably defined permutation
matrix \cite{Pailloux2011};
\item $\alpha\in\mathbb{C}$ is an unknown deterministic factor accounting
for both channel reflectivity and channel propagation effects;
\item $\bm{n}_{0}\sim\mathcal{CN}_{N}(\bm{0}_{N},\bm{M}_{0})$ and $\bm{n}_{0k}\sim\mathcal{CN}_{N}(\bm{0}_{N},\gamma\bm{M}_{0})$,
$k=1,\ldots,K$, where the positive definite covariance matrix $\bm{M}_{0}\in\{\bm{R}\in\mathbb{H}^{N\times N}\,:\,\bm{R=}\bm{J}\bm{R}^{*}\bm{J}\}$
and the scaling factor $\gamma\in\mathbb{R}^{+}$ are both \emph{unknown}
deterministic quantities (the latter assumption determines a PHE).
\end{itemize}
The model in Eq. (\ref{eq: Original hypothesis testing problem})
can be recast in the more advantageous \emph{canonical form}, as shown
in \cite{A.DeMaio2015}. Thus, without loss of generality, we can
express the problem as:
\begin{gather}
\begin{cases}
\mathcal{H}_{0}: & \begin{cases}
\bm{z}_{1}=\bm{n}_{1},\;\bm{z}_{2}=\bm{n}_{2},\\
\bm{z}_{1k}=\bm{n}_{1k},\:\bm{z}_{2k}=\bm{n}_{2k},\quad k=1,\ldots K,
\end{cases}\\
\mathcal{H}_{1}: & \begin{cases}
\bm{z}_{1}=\alpha_{1}\,\bm{e}_{1}+\bm{n}_{1},\;\bm{z}_{2}=\alpha_{2}\,\bm{e}_{1}+\bm{n}_{2},\\
\bm{z}_{1k}=\bm{n}_{1k},\;\bm{z}_{2k}=\bm{n}_{2k},\quad k=1,\ldots,K,
\end{cases}
\end{cases},\label{eq: canonical problem}
\end{gather}
where we have adopted the notation $\bm{z}_{1}\triangleq\bm{V}\,\Re\{\bm{T}\bm{r}\}\in\mathbb{R}^{N\times1}$,
$\bm{z}_{2}\triangleq\bm{V}\,\Im\{\bm{T}\bm{r}\}\in\mathbb{R}^{N\times1}$,
$\bm{z}_{1k}\triangleq\bm{V}\,\Re\{\bm{T}\bm{r}_{k}\}\in\mathbb{R}^{N\times1}$
and $\bm{z}_{2,k}\triangleq\bm{V}\,\Im\{\bm{T}\bm{r}_{k}\}\in\mathbb{R}^{N\times1}$,
$k=1,\ldots K$, for the transformed primary and secondary data, respectively.
We recall that the unitary matrix $\bm{T}\in\mathbb{C}^{N\times N}$
(whose definition is provided in \cite{Pailloux2011}) and the orthogonal
matrix $\bm{V}\in\mathbb{R}^{N\times N}$ (any orthogonal matrix whose
first row is aligned to $\bm{T}\bm{s}$) are needed to obtain an equivalent
real-valued representation of the persymmetric model and rotate the
space into the canonical basis, respectively. Additionally, $\alpha_{1}\triangleq\Re\{\alpha\}$
and $\alpha_{2}\triangleq\Im\{\alpha\}$ denote the unknown deterministic
coefficients accounting for the useful signal (collected in the vector
$\bm{\alpha}\triangleq\begin{bmatrix}\alpha_{1} & \alpha_{2}\end{bmatrix}^{T}$),
whereas $\bm{e}_{1}\triangleq\begin{bmatrix}1 & 0 & \cdots & 0\end{bmatrix}^{T}\in\mathbb{R}^{N\times1}$
denotes the steering vector in canonical representation. With reference
to the disturbance, we have employed the analogous definitions $\bm{n}_{1}\triangleq\bm{V}\,\Re\{\bm{T}\bm{n}_{0}\}$,
$\bm{n}_{2}\triangleq\bm{V}\,\Im\{\bm{T}\bm{n}_{0}\}$, $\bm{n}_{1k}\triangleq\bm{V}\,\Re\{\bm{T}\bm{n}_{0k}\}$
and $\bm{n}_{2,k}\triangleq\bm{V}\,\Im\{\bm{T}\bm{n}_{0k}\}$, $k=1,\ldots K$,
with $\bm{n}_{i}\sim\mathcal{N}_{N}(\bm{0}_{N},\bm{M})$ and $\bm{n}_{i,k}\sim\mathcal{N}_{N}(\bm{0}_{N},\gamma\,\bm{M})$,
$i=1,2$, $k=1,\ldots K$, where $\bm{M}\triangleq(1/2)\,\bm{V}\bm{T}\bm{M}_{0}\bm{T}^{\dagger}\bm{V}^{T}$
represents the transformed (real-valued) covariance matrix of the
primary data.

Before proceeding further, we collect all the secondary data in $\bm{Z}_{s}\triangleq\begin{bmatrix}\bm{z}_{11} & \cdots & \bm{z}_{1K} & \bm{z}_{21} & \cdots & \bm{z}_{2K}\end{bmatrix}\in\mathbb{R}^{N\times2K}$
and give the following preliminary definitions\footnote{These definitions will be thoroughly exploited in the derivation of
the MIS in Sec. \ref{sec: MIS}.}:
\begin{equation}
\bm{Z}_{p}\triangleq\begin{bmatrix}\bm{z}_{1} & \bm{z}_{2}\end{bmatrix}=\begin{bmatrix}\bm{z}_{1p}\\
\bm{Z}_{2p}
\end{bmatrix},\;\bm{S}\triangleq\bm{Z}_{s}\bm{Z}_{s}^{T}=\begin{bmatrix}s_{11} & \bm{s}_{12}\\
\bm{s}_{21} & \bm{S}_{22}
\end{bmatrix},\label{eq: Z_c S_c (block definition)}
\end{equation}
where $\bm{z}_{1p}\in\mathbb{R}^{1\times2}$ (i.e., a row vector),
$\bm{Z}_{2p}\in\mathbb{R}^{(N-1)\times2}$, $s_{11}\in\mathbb{R}$
(i.e., a scalar), $\bm{s}_{12}\in\mathbb{R}^{1\times(N-1)}$ (i.e.,
a row vector), $\bm{s}_{21}\in\mathbb{R}^{(N-1)\times1}$ and $\bm{S}_{22}\in\mathbb{R}^{(N-1)\times(N-1)}$
respectively. Furthermore, it is easily shown that $\bm{Z}_{p}|\mathcal{H}_{1}\sim\mathcal{N}_{N\times2}(\bm{e}_{1}\bm{\alpha}^{T},\bm{I}_{2},\bm{M})$
(resp. $\bm{Z}_{p}|\mathcal{H}_{0}\sim\mathcal{N}_{N\times2}(\bm{0}_{N\times2},\bm{I}_{2},\bm{M})$),
whereas $\bm{Z}_{s}\sim\mathcal{N}_{N\times2K}(\bm{0}_{N\times2K},\bm{I}_{2K},\gamma\bm{M})$.

In this letter we will consider decision rules which declare $\mathcal{H}_{1}$
(resp. $\mathcal{H}_{0}$) if $\Phi\left(\bm{Z}_{p},\bm{S}\right)\geq\eta$
(resp. $\Phi\left(\bm{Z}_{p},\bm{S}\right)<\eta$), where $\Phi(\cdot)\,:\,\mathbb{R}^{N\times2}\times\mathbb{S}^{N\times N}\rightarrow\mathbb{R}$
indicates the generic form of a decision function based on the sufficient
statistic\footnote{In fact, Fisher-Neyman factorization theorem ensures that the optimal
decision from $\{\bm{Z}_{p},\bm{S}\}$ is tantamount to deciding from
raw data \cite{Muirhead2009}.} $(\bm{Z}_{p},\bm{S})$ and $\eta$ denotes the threshold set to ensure
a desired false-alarm probability ($P_{fa}$).

\section{Maximal Invariant Statistic\label{sec: MIS}}

In what follows, we will search for functions of data sharing invariance
w.r.t. those parameters (namely, the nuisance parameters $\bm{M}$
and $\gamma$) which are irrelevant for the specific decision problem.
To this end, we resort to the so-called ``Principle of Invariance''
\cite{Lehmann2006}, whose main idea consists in finding transformations
that properly cluster data without altering: ($i$) the formal structure
of the hypothesis testing problem given by $\mathcal{H}_{0}:\left\Vert \bm{\alpha}\right\Vert =0$,
$\mathcal{H}_{1}:\left\Vert \bm{\alpha}\right\Vert >0$; ($ii$) the
Gaussian assumption for the received data under each hypothesis; ($iii$)
the real symmetric structure of the covariance matrix and the useful
signal subspace. Therefore, next subsection is devoted to the definition
of a suitable group which fulfills the above requirements.

\subsection{Desired invariance properties \label{sub: Desired Invariance Properties}}

First, without loss of generality, we will consider transformations
acting directly on the sufficient statistic $\{\bm{Z}_{p},\bm{S}\}$.
Then, denote by $\mathcal{G}_{N}$ the linear group of (real-valued)
$N\times N$ non-singular matrices having the peculiar structure
\begin{gather}
\bm{G}\triangleq\begin{bmatrix}g_{11} & \bm{g}_{12}\\
\bm{0} & \bm{G}_{22}
\end{bmatrix}\,,
\end{gather}
where $g_{11}\neq0$ and $\mathrm{det}(\bm{G}_{22})\neq0$. Also,
let $\mathcal{O}_{2}$ represent the group of $2\times2$ orthogonal
matrices (with generic element denoted with $\bm{U}$) and consider
the set $\mathbb{R}^{+}$ (with generic element denoted with $\varphi$),
along with the composition operator ``$\circ$'' defined as
\begin{equation}
(\bm{G}_{a},\bm{U}_{a},\varphi_{a})\circ(\bm{G}_{b},\bm{U}_{b},\varphi_{b})=(\bm{G}_{b}\bm{G}_{a},\bm{U}_{a}\bm{U}_{b},\varphi_{a}\varphi_{b})\,.\label{eq:  group composition operation}
\end{equation}
The sets and the composition operator are here represented compactly
as $\mathcal{L}\triangleq(\mathcal{G}_{N}\times\mathcal{O}_{2}\times\mathbb{R}^{+},\circ)$,
in a \emph{group}\footnote{Indeed $\mathcal{L}$ satisfies the following elementary axioms: ($i$)
it is \emph{closed} w.r.t. the operation ``$\circ$'', ($ii$) it
satisfies the associative property and ($iii$) there exist both the
identity and the inverse elements.} form. The group $\mathcal{L}$ has the fundamental property of leaving
the hypothesis testing problem in Eq. (\ref{eq: canonical problem})
\emph{invariant} under the action $\ell(\cdot,\cdot)$, defined as
follows: 
\begin{equation}
\ell(\bm{Z}_{p},\bm{S})=\left(\bm{G}\bm{Z}_{p}\bm{U},\varphi\,\bm{G}\,\bm{S}\,\bm{G}^{T}\right)\quad\forall(\bm{G},\bm{U},\varphi)\in\mathcal{L}\,.\label{eq: elementary action}
\end{equation}
The proof of the aforementioned statement is straightforward and is
omitted due to lack of space. Such property implies that $\mathcal{L}$
preserves the family of distributions (i.e., $\bm{G}\bm{Z}_{p}\bm{U}$
and $\varphi\,\bm{G}\,\bm{S}\,\bm{G}^{T}$ remain Gaussian- and Wishart-distributed
as $\bm{Z}_{p}$ and $\bm{S}$, respectively), as well as the structure
of the hypothesis testing problem considered. Additionally, $\mathcal{L}$
is chosen to include those transformations which are of practical
interest, as they allow claiming the CFAR property (w.r.t. $\bm{M}$
and $\gamma$) as a byproduct of the invariance.

\subsection{Derivation of the MIS\label{sub: MIS Derivation}}

In Sec. \ref{sub: Desired Invariance Properties}, we have identified
a group $\mathcal{L}$ which leaves the problem under investigation
unaltered. As a consequence, we are thus reasonably motivated to search
for decision rules that are invariant under $\mathcal{L}$. To this
end, we invoke the Principle of Invariance because it allows to construct
statistics (viz. the MISs) that organize data into distinguishable
equivalence classes (named \emph{orbits}). Every invariant test (w.r.t.
$\mathcal{L}$) can be written as a function of the corresponding
maximal invariant \cite{Scharf1991}.

The MIS (w.r.t. $\mathcal{L}$) satisfies both the properties:
\begin{align}
(a)\:\bm{T}(\bm{Z}_{p},\bm{S}) & =\bm{T}(\ell(\bm{Z}_{p},\bm{S})),\;\forall\ell\:\mathrm{action\:of}\:\mathcal{L},\\
(b)\:\bm{T}(\bm{Z}_{p},\bm{S}) & =\bm{T}(\bar{\bm{Z}}_{p},\bar{\bm{S}})\nonumber \\
 & \Rightarrow\exists\:\ell\:\mathrm{action\:of}\:\mathcal{L}:\,(\bm{Z}_{p},\bm{S})=\ell(\bar{\bm{Z}}_{p},\bar{\bm{S}})\,.
\end{align}
Conditions ($a$) and ($b$) correspond to \emph{invariance} and \emph{maximality}
properties, respectively. The explicit expression of the MIS for the
problem at hand is provided in the following proposition.
\begin{prop}
A MIS w.r.t. $\mathcal{L}$ for the problem in Eq. (\ref{eq: canonical problem})
is given by the vector:\label{prop: Maximal Invariant Statistic}
\begin{gather}
\bm{t}(\bm{Z}_{p},\bm{S})\triangleq\begin{bmatrix}t_{1} & t_{2} & t_{3}\end{bmatrix}^{T}=\begin{bmatrix}\lambda_{1}/\lambda_{4} & \lambda_{2}/\lambda_{4} & \lambda_{3}/\lambda_{4}\end{bmatrix}^{T},\label{eq: MIS_final}
\end{gather}
where $\lambda_{i}$, $i=1,2$ and $\lambda_{j}$, $j=3,4$, are the
eigenvalues of $\bm{\Psi}_{0}\triangleq\bm{Z}_{p}^{T}\bm{S}^{-1}\bm{Z}_{p}$
and $\bm{\Psi}_{1}\triangleq\bm{Z}_{2p}^{T}\bm{S}_{22}^{-1}\bm{Z}_{2p}$,
respectively.\end{prop}
\begin{IEEEproof}
The proof is given in Appendix \ref{sec: Appendix_ MIS derivation}.
\end{IEEEproof}
Some important remarks are now in order. The MIS is given by a 3-D
vector, where the third component ($t_{3}$) represents an \emph{ancillary
part, }that is, its distribution does not depend on the hypothesis
in force. Furthermore, exploiting \cite[Thm. 6.2.1]{Lehmann2006},
every invariant statistic may be written as a function of Eq.~(\ref{eq: MIS_final}).
Therefore, it follows that every invariant test is CFAR.

Finally, we conclude the section with a discussion on the maximal
invariant induced in the parameter space \cite{Lehmann2006}, representing
the reduced set of unknown parameters on which the hypothesis testing
in the invariant domain depends. To this end, we observe that the
pdf of $\bm{t}(\bm{Z}_{p},\bm{S})$ does not depend on $\gamma$,
because of the normalization by $\lambda_{4}$. Thus, following the
lead of \cite{A.DeMaio2015}, it can be shown that the induced maximal
invariant corresponds to the Signal-to-Interference plus Noise Ratio
(SINR) $\left\Vert \bm{\alpha}\right\Vert ^{2}\,\bm{e}_{1}^{T}\,\bm{M}^{-1}\,\bm{e}_{1}$,
which is the same as the case of the HE. As a result, when the hypothesis
$\mathcal{H}_{0}$ is in force, the $\mathrm{SINR}$ equals zero and
thus the pdf of $\bm{t}(\bm{Z}_{p},\bm{S})$ does not depend on any
unknown parameter. Therefore every function of the MIS satisfies the
CFAR property.

\subsection{Detectors Design vs. the MIS\label{sub: MIS statistical distribution}}

This subsection is devoted to the design of detectors based on well-founded
criteria. Accordingly, we will concentrate on the derivation of the
well-known GLRT (including its two-step version), Rao, and Wald tests
\cite{Kay1998}. 

Before proceeding, we first report the explicit expressions of the
Maximum Likelihood (ML) estimates of the scale parameters under both
hypotheses ($\hat{\gamma}_{i}$, $i=0,1$):
\begin{gather}
\widehat{\gamma}_{i}\triangleq\frac{\beta_{i}-(K+1-N)\mathrm{Tr}[\bm{\Psi}_{i}]}{2(2K+2-N)\,\mathrm{det}\left(\bm{\Psi}_{i}\right)},\label{eq: gamma_i def}\\
\beta_{i}\triangleq\sqrt{\mathrm{Tr}[\bm{\Psi}_{i}]^{2}(K+1-N)^{2}+4N(2K+2-N)\mathrm{det}(\bm{\Psi}_{i})},
\end{gather}
The  definition in (\ref{eq: gamma_i def}) will be exploited to provide
compact expressions of the following detectors. Indeed, the GLR is
given by \cite{Casillo2007}:
\begin{equation}
t_{\mathrm{glr}}\triangleq\frac{\widehat{\gamma}_{0}^{-\frac{N}{K+1}}\mathrm{det}\left(\bm{I}_{2}+\widehat{\gamma}_{0}\bm{\Psi}_{0}\right)}{\widehat{\gamma}_{1}^{-\frac{N}{K+1}}\mathrm{det}\left(\bm{I}_{2}+\widehat{\gamma}_{1}\bm{\Psi}_{1}\right)}\,,\label{eq: t_glr}
\end{equation}
while the 2S-GLR (also referred in the literature to as Per-ACE) is
\cite{Gao2014}:
\begin{equation}
t_{\mathrm{2s-glr}}\triangleq\mathrm{Tr}[\bm{\Psi}_{0}]\,/\,\mathrm{Tr}[\bm{\Psi}_{1}]\,.
\end{equation}
Differently, the Rao statistic is given by \cite{Hao2012a}:
\begin{gather}
t_{\mathrm{rao}}\triangleq\frac{\hat{\gamma}_{0}\mathrm{Tr}\left[(\bm{\Psi}_{0}-\bm{\Psi}_{1})\left(\bm{I}_{2}+\widehat{\gamma}_{0}\bm{\Psi}_{0}\right)^{-2}\right]}{1-\widehat{\gamma}_{0}\mathrm{Tr}\left[(\bm{\Psi}_{0}-\bm{\Psi}_{1})\left(\bm{I}_{2}+\widehat{\gamma}_{0}\bm{\Psi}_{0}\right)^{-1}\right]}\,.\label{eq: Rao statistic (MIS ready)}
\end{gather}
Finally, the Wald statistic is \cite{Hao2012a}:
\begin{eqnarray}
t_{\mathrm{wald}} & \triangleq & \widehat{\gamma}_{1}\left\{ \mathrm{Tr}[\bm{\Psi}_{0}]-\mathrm{Tr}[\bm{\Psi}_{1}]\right\} \,.\label{eq: Wald statistic}
\end{eqnarray}
We now show that the aforementioned statistics are all functions of
the data \emph{solely through the MIS} $\bm{t}(\bm{Z}_{p},\bm{S})$
(viz. the corresponding tests ensure a CFAR). The detailed proof of
this claim is provided as supplementary material for this letter.
First, it is proved that $t_{\mathrm{glr}}$ in (\ref{eq: t_glr})
can be rewritten as:
\begin{align}
t_{\mathrm{glr}}= & \left\{ \frac{t_{3}\,g_{\gamma}(t_{1}/t_{2})}{t_{1}\,g_{\gamma}(t_{3})}\right\} ^{-\frac{N}{K+1}}\nonumber \\
 & \;\times\frac{\left[1+g_{\gamma}(t_{1}/t_{2})\right]\left[1+(t_{2}/t_{1})\,g_{\gamma}(t_{1}/t_{2})\right]}{\left[1+g_{\gamma}(t_{3})\right]\left[1+(1/t_{3})\,g_{\gamma}(t_{3})\right]}\,,\label{eq: t_glr (MIS)}
\end{align}
where $g_{\gamma}(t_{1}/t_{2})$ (resp. $g_{\gamma}(t_{3})$) denotes
a suitably defined auxiliary function which depends on the data solely
through the ratio $t_{1}/t_{2}$ (resp. through $t_{3}$). Secondly,
the 2S-GLR (Per-ACE) is rewritten as:
\begin{equation}
t_{\mathrm{2s-glr}}=(t_{1}+t_{2})\,/\,(1+t_{3})\,.\label{eq: t_2sglr (MIS)}
\end{equation}
Differently, proving this property for Rao statistic is much more
involved and it is based on showing that the terms $\widehat{\gamma}_{0}\mathrm{Tr}\left[(\bm{\Psi}_{0}-\bm{\Psi}_{1})\left(\bm{I}_{2}+\widehat{\gamma}_{0}\bm{\Psi}_{0}\right)^{-2}\right]$
and $\widehat{\gamma}_{0}\mathrm{Tr}\left[(\bm{\Psi}_{0}-\bm{\Psi}_{1})\left(\bm{I}_{2}+\widehat{\gamma}_{0}\bm{\Psi}_{0}\right)^{-1}\right]$
are \emph{both invariant.} Finally, Wald statistic can be rewritten
as:
\begin{eqnarray}
t_{\mathrm{wald}} & = & g_{\gamma}(t_{3})\left\{ (t_{1}/t_{3})+(t_{2}/t_{3})-\left(1+1/t_{3}\right)\right\} ,\label{eq: Wald (MIS)}
\end{eqnarray}
which proves its invariance.

\section{Conclusions\label{sec: Conclusions}}

In this letter we studied adaptive detection of a point-like target
in the presence of PHE with persymmetric-structured covariance by
resorting to statistical theory of invariance. After obtaining the
group of transformations leaving the hypothesis testing problem invariant
(thus enforcing at the design stage the CFAR property), a MIS was
derived for the aforementioned group. It was found that the MIS for
the problem at hand is a 3-D vector, with the latter component being
an ancillary term. Subsequently, we focused on the derivation of (2S-)
GLR, Rao and Wald statistics for the considered problem. Remarkably,
all the aforementioned statistics were shown to be functions of the
data solely through the MIS and, hence, to share the invariance property
w.r.t. $\mathcal{L}$. As a consequence, they ensure a CFAR w.r.t.
the unknown parameters of the disturbance.

\section{Acknowledgement}

The authors would like to thank Prof. A. De Maio, at University of
Naples ``Federico II'', for suggesting the topic and for the interesting
discussions towards the solution of the problem.

\appendices{}

\section{Proof of Proposition \ref{prop: Maximal Invariant Statistic}\label{sec: Appendix_ MIS derivation}}

The proof is based on the key observation that the action $\ell(\cdot,\cdot)$
(cf. Eq. (\ref{eq: elementary action})) can be re-interpreted as
the \emph{sequential} application of the following sub-actions:
\begin{gather}
\ell_{1}(\bm{Z}_{p},\bm{S})=(\bm{G}\,\bm{Z}_{p}\bm{U},\bm{G}\,\bm{S}\,\bm{G}^{T})\quad\forall(\bm{G},\bm{U})\in\mathcal{L}_{1}\nonumber \\
\ell_{2}(\bm{Z}_{p},\bm{S})=\left(\bm{Z}_{p},\varphi\bm{S}\right)\quad\forall\varphi\in\mathcal{L}_{2},
\end{gather}
where $\mathcal{L}_{1}\triangleq\{\mathcal{G}_{N}\times\mathcal{O}_{2},``\circ"\}$
and $\mathcal{L}_{2}\triangleq\{\mathbb{R}^{+},``\times"\}$ (i.e.,
the composition operator for $\mathcal{L}_{2}$ simply corresponds
to the product). Then, it is recognized that the MIS for the sub-action
$\ell_{1}(\cdot,\cdot)$ has been already obtained in \cite{A.DeMaio2015},
as the former represents the relevant action enforcing desirable invariances
in a homogeneous background (viz. $\gamma=1$). Such statistic is
$4$-D and given by $\bm{t}_{\mathrm{HE}}(\bm{Z}_{p},\bm{S})\triangleq\begin{bmatrix}\lambda_{1} & \lambda_{2} & \lambda_{3} & \lambda_{4}\end{bmatrix}^{T}$,
where $\lambda_{1}\geq\lambda_{2}$ are the two eigenvalues of $\bm{Z}_{p}^{T}\bm{S}^{-1}\bm{Z}_{p}$
and $\lambda_{3}\geq$ $\lambda_{4}$ denote the two eigenvalues of
$\bm{Z}_{2p}^{T}\bm{S}_{22}^{-1}\bm{Z}_{2p}$, respectively. Now,
define the action $\ell_{2}^{\star}(\cdot)$ acting on the couple
of positive-valued scalars $a_{i}$, collected in the vector $\bm{a}\triangleq\begin{bmatrix}a_{1} & a_{2} & a_{3} & a_{4}\end{bmatrix}^{T}$
(with $a_{i}$ corresponding to $\lambda_{i}$) as:
\begin{equation}
\ell_{2}^{\star}(\bm{a})=\varphi^{-1}\bm{a}\quad\forall\,\varphi\in\mathcal{L}_{2}.\label{eq: l_2,b adjoint}
\end{equation}
It is not difficult to show that a MIS for the elementary operation
$\ell_{2}^{\star}(\cdot)$ in Eq. (\ref{eq: l_2,b adjoint}) is given
by $\bm{t}_{2}(\bm{a})\triangleq\begin{bmatrix}\frac{a_{1}}{a_{4}} & \frac{a_{2}}{a_{4}} & \frac{a_{3}}{a_{4}}\end{bmatrix}^{T}$.
This is clearly achieved by verifying that both \emph{invariance}
and \emph{maximality} properties \cite{Lehmann2006} hold for $\bm{t}_{2}(\cdot)$.
Indeed \emph{invariance} follows from $\bm{t}_{2}(\varphi^{-1}\bm{a})=\begin{bmatrix}\frac{\varphi^{-1}a_{1}}{\varphi^{-1}a_{4}} & \frac{\varphi^{-1}a_{2}}{\varphi^{-1}a_{4}} & \frac{\varphi^{-1}a_{3}}{\varphi^{-1}a_{4}}\end{bmatrix}^{T}=\bm{t}_{2}(\bm{a})$,
while \emph{maximality} can be proved as follows. Suppose that $\bm{t}_{2}(\bm{a})=\bm{t}_{2}(\bar{\bm{a}})$
holds, which implies:
\begin{equation}
\bar{a}_{i}=(\bar{a}_{4}/a_{4})\,a_{i},\quad i=1,2,3.
\end{equation}
Then there exists a $\varphi\in\mathcal{L}_{2}$, equal to $\varphi=\frac{a_{4}}{\bar{a}_{4}}$,
which ensures $\left(\varphi^{-1}\bm{a}\right)=\bar{\bm{a}}$. This
demonstrates that $\bm{t}_{2}(\bm{a})$ is a MIS for $\ell_{2}^{\star}(\cdot)$.
Additionally, we notice that 
\begin{gather}
\bm{t}_{\mathrm{HE}}(\bar{\bm{Z}}_{p},\bar{\bm{S}})=\bm{t}_{\mathrm{HE}}(\bm{Z}_{p},\bm{S})\Rightarrow\nonumber \\
\bm{t}_{\mathrm{HE}}(\bar{\bm{Z}}_{p},\varphi\bar{\bm{S}})=\bm{t}_{\mathrm{HE}}(\bm{Z}_{p}\,,\varphi\bm{S})\,,\:\forall\varphi\in\mathcal{L}_{2},
\end{gather}
since $\bm{t}_{\mathrm{HE}}(\bm{Z}_{p},\varphi\bm{S})=\frac{1}{\varphi}\bm{t}_{\mathrm{HE}}(\bm{Z}_{p},\bm{S})$
holds for the problem at hand. Therefore, exploiting \cite[p. 217, Thm. 6.2.2]{Lehmann2006},
it follows that a MIS for the action $\ell(\cdot,\cdot)$ is given
by the composite function $\bm{t}(\bm{Z}_{p},\bm{S})\triangleq\bm{t}_{2}(\bm{t}_{\mathrm{HE}}(\bm{Z}_{p},\bm{S}))=\begin{bmatrix}\lambda_{1}/\lambda_{4} & \lambda_{2}/\lambda_{4} & \lambda_{3}/\lambda_{4}\end{bmatrix}^{T}$.
This concludes our proof.

\bibliographystyle{IEEEtran}
\bibliography{IEEEabrv,bib_adapt_det}

\end{document}